\documentclass[showpacs,aps,prb,superscriptaddress,twocolumn]{revtex4}

\usepackage{amsmath}
\usepackage{amssymb}
\usepackage{amsbsy}
\usepackage{makeidx}
\usepackage{epsf}
\usepackage{graphicx}
\usepackage{amsfonts}
\usepackage{subfigure}

\def\intd{\int_{-\infty}^\infty}

\begin{document}
\title{Localized mode interactions in 0-$\pi$ Josephson junctions}
\author{Hadi Susanto}
\affiliation{School of Mathematical Sciences, University of Nottingham, University Park, Nottingham, NG7 2RD, UK}
\author{Gianne Derks}
\affiliation{Department of Mathematics,
University of Surrey, Guildford, Surrey, GU2 7XH, UK}

\begin{abstract}
  A long Josephson junction containing regions with a phase shift of
  $\pi$ is considered. By exploiting the defect modes due to the
  discontinuities present in the system, it is shown that Josephson
  junctions with phase-shift can be an ideal
  setting for studying localized mode interactions. A phase-shift
  configuration acting as a double-well potential is
  considered and shown to admit mode tunnelings between the
  wells. When the phase-shift
configuration
is periodic, it is shown that localized excitations forming bright and
dark solitons can be created. Multi-mode approximations are derived confirming the numerical results.
\end{abstract}


\maketitle

\newpage

\emph{Introduction}. A Josephson junction is a system consisting of
two layers of superconductors separated by a nonsuperconducting
barrier. Electrons forming so-called Cooper pairs can tunnel across
the resistive barrier even when there is no applied voltage
difference. Theoretically predicted by Josephson\cite{jose62} and
first observed experimentally in Ref.\ \onlinecite{ande63}, the only
requirement for the occurrence of Josephson tunneling is a weak
coupling of the wave functions of the two superconductors. The
supercurrent ($I_s$) is proportional to the sine of the electron
phase-difference across the insulator ($u$), i.e.\ $I_s\sim\sin{u}$.

Bulaevskii et al.\cite{bula77,bula78}
proposed that
a shift of $\pi$ can be introduced in the phase difference $u$ of a
Josephson junction by installing magnetic impurities, which has been
confirmed recently\cite{vavr06}. Present technological advances can
also impose a $\pi$-phase-shift in a long Josephson junction using
various means, including multilayer junctions with controlled
thicknesses over the insulating barrier\cite{ryaz01,base99}, pairs of
current injectors \cite{gold04}, and junctions with unconventional
order parameter symmetry\cite{tsue00,hilg03,guma07}.

New phenomena may occur when a junction with phase-shifts is connected
to a normal junction, i.e.\ 0-$\pi$ Josephson junctions. These include
the presence of a half magnetic flux quantum induced by spontaneously
created supercurrent circulating in a loop. Such unique
characteristics offer promising future device applications, such as
novel circuits for information storage and processing in both
classical and quantum limits\cite{gold05} and artificial crystals for
simulating and studying energy levels and band structures in large
systems of spins\cite{susa05} (see also Ref.\ \onlinecite{hilg08} and
references therein). Here, we demonstrate that Josephson junctions
with phase-shifts
is an ideal setting
for showcasing many interesting features of localized mode
interactions. Arguably the dynamics of the superconductor
phase-difference can be seen to be analogues to that of atomic wave
functions of Bose-Einstein condensates
(BEC)\cite{bose24,eins25,ande95,davi95,brad97} in an external
potential. In particular, we consider Josephson junctions with
phase-shift configurations acting as a double well and a periodic
potential.

The interesting phenomenon of mode tunneling in BECs in a double well
potential was predicted by Smerzi et al.\ \cite{smer97,giov00},
followed by
experimental observations\cite{albi05,lebl10}. Here, we
will show that defect modes due to the presence of
phase-discontinuities in $0$-$\pi$ Josephson junctions can be exploited
to observe a similar mode tunneling, which can also be viewed as Rabi
oscillations of two interacting modes.\cite{ostr00} Periodic defects
exhibiting mode self-trapping analogues to BECs in optical
lattices\cite{mors06} will also be discussed.

\emph{Double-well potential}. A 0-$\pi$ Josephson junction with the superconductor phase difference $u$ at position $x$ and time $t$ is described by the sine-Gordon equation
\begin{equation}
u_{tt}-u_{xx}=-\theta(x)\sin{u},
\label{sg}
\end{equation}
where $x$ and $t$ have been normalized to the Josephson penetration
depth $\lambda_J$ and the inverse plasma frequency $\omega_p^{-1}$,
respectively. The function $\theta(x)$ is piecewise constant
representing the presence of $\pi$ junctions.
A double well potential with two $\pi$-junctions of length $a$ separated by a
$0$-junction with length $2L$ is described by
\begin{equation}
\theta(x)=\left\{
\begin{array}{rl}
-1,& \quad L<|x|<L+a,\\
1,& \quad \rm{elsewhere}.
\end{array}
\right.
\end{equation}
%
At the points of discontinuities, the boundary conditions are
\begin{equation}
\begin{array}{rcl}
\displaystyle\lim_{x\to\pm\{L,L+a\}^+}u(x,t)&=&
\displaystyle\lim_{x\to\pm\{L,L+a\}^-}u(x,t);\\
\displaystyle\lim_{x\to\pm\{L,L+a\}^+}u_x(x,t)&=&
\displaystyle\lim_{x\to\pm\{L,L+a\}^-}u_x(x,t).
\end{array}
\label{bc}
\end{equation}

If $\phi$ solves (\ref{sg}), the linear stability of the solution can be
analyzed by substituting the spectral ansatz $u=\phi+v(x)e^{\lambda
  t}$ and linearizing about $||v||_\infty$ small to yield the
eigenvalue problem $v_{xx}-\lambda^2v=\theta\cos(\phi)v.$

Equation (\ref{sg}) has two constant solutions (mod $2\pi$),
$u\equiv0$ and $u\equiv\pi$. The solution $u\equiv\pi$ has unstable
continuous spectrum and hence is always unstable.\cite{susa03} The
solution $u=0$ has stable continuous spectrum $\lambda^2<-1$ and the
discrete spectrum (eigenvalues) can be calculated
analytically.\cite{susa03} Indeed, the largest two eigenvalues
$\Lambda_\pm$ of $u=0$ solve the equation
\begin{equation}
  \label{eq:ev+-}
  \frac{\sqrt{1-\Lambda_\pm^2}}{\tan\left(\sqrt{1-\Lambda_\pm^2}a\right)}
  +\Lambda_\pm \mp e^{-2\sqrt{1+\Lambda_\pm}L} =0.
\end{equation}
The corresponding eigenmodes
$\Phi_{\pm}(x)$ of $\Lambda_\pm$ are\cite{susa03}
\begin{equation}
\Phi_\pm = \left\{
\begin{array}{ll}
e^{-\sqrt{1+\Lambda_\pm}(x-L-a)},&\,x>L+a; \\
\cos(\sqrt{1-\Lambda_\pm}(x-L-a)) \\
{}+ C\,\sin(\sqrt{1-\Lambda_\pm}(x-L-a)),&\,L<x<L+a; \\
K_+\,\cosh(\sqrt{1+\Lambda_\pm}x) \\
{}+ K_-\,\sinh(\sqrt{1+\Lambda_\pm}x)
,&\,0<x<L,
\end{array}
\right.
\label{phipm}
\end{equation}
where $C=-\sqrt{\frac{1+\Lambda_\pm}{1-\Lambda_\pm}}$ and $K_\pm =
\frac{2e^{-\sqrt{1+\Lambda_\pm}L}\sin(\sqrt{1-\Lambda_\pm}a)}
{\sqrt{1-\Lambda_\pm^2}}$. As the linearisation operator is a
Sturm-Liouville operator and even in~$x$, the eigenmodes are simple
and the eigenfunction $\Phi_+$ is an even function and $\Phi_-$ is
odd. The two eigenvalues $\Lambda_\pm$ are depicted as a function of
$a$ for fixed $L=2$ (to the left of the vertical dashed line).

It is clear that $u=0$ has a stability window. The change of stability
occurs at a critical distance $a_c$ when the critical eigenvalue
crosses the horizontal axis $\Lambda=0$. Using the expression
in~\eqref{eq:ev+-}, it follows that the critical length $a_c(L)$
is
$a_c=\arctan(e^{2L})$.

\begin{figure}[tbhp]
\centering
\includegraphics[width=6cm,angle=0,clip]{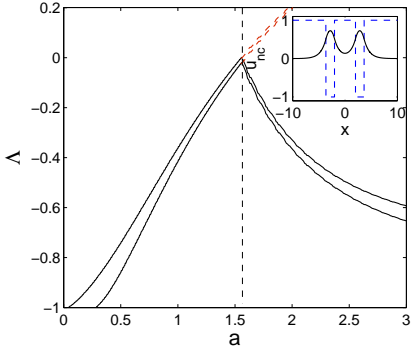}
\caption{The eigenvalues of the ground state as a
  function of $a$ for $L=2$. The dashed vertical line indicates the bifurcation point of the non uniform ground state, where on the left and on the right of the vertical line $u=0$ is stable and unstable, respectively. The dashed lines show the eigenvalues of the uniform solution in its instability region. The inset presents a ground state when $L=2$ and $a=1.65$. 
  }
\label{stabof0}
\end{figure}

In the instability region, a non uniform time-independent sign-definite ground state $\pm u_{nc}(x)$ bifurcates. Its expression can be written in terms of Jacobian elliptic functions \cite{susa03,zenc04} as
\begin{equation}
u_{nc}=
\left\{
  \begin{array}{rl}
   4\arctan(e^{-x+x_0}), &\, x>L+a;\\
   2\arcsin(m_1\mathop{\rm sn}(x-x_1,m_1)), &\, L<x<L+a;\\
   \pi+2\arcsin(m_2\mathop{\rm sn}(x-x_2,m_2)),  &\, 0<x<L.
  \end{array}
\right.
\label{nc}
\end{equation}
The parameters $m_1$ and $m_2$ are linked to the lengths $a$ and $L$
by
\[
\begin{array}{rcl}
a&=&2K(m_1)-\mathop{\rm am}^{-1}(\pi/4,m_1)\\[1mm]
&&-\mathop{\rm am}^{-1}(\arcsin(\sqrt{2(1+m_1^2-m_2^2)}/2m_1),m_1)\\[1mm]
L&=&K(m_2)-\mathop{\rm am}^{-1}(\arcsin(\sqrt{2(1+m_2^2-m_1^2)}/2m_2),m_2)
\end{array}
\]
The translations $x_i$ are determined by the boundary
conditions~\eqref{bc}.  The non-uniform ground state and its
eigenvalues are presented in Fig.\ \ref{stabof0}
(to the right of the vertical dashed line).

\begin{figure}[tbhp]
\centering
\subfigure[]{\includegraphics[width=4cm,angle=0,clip]{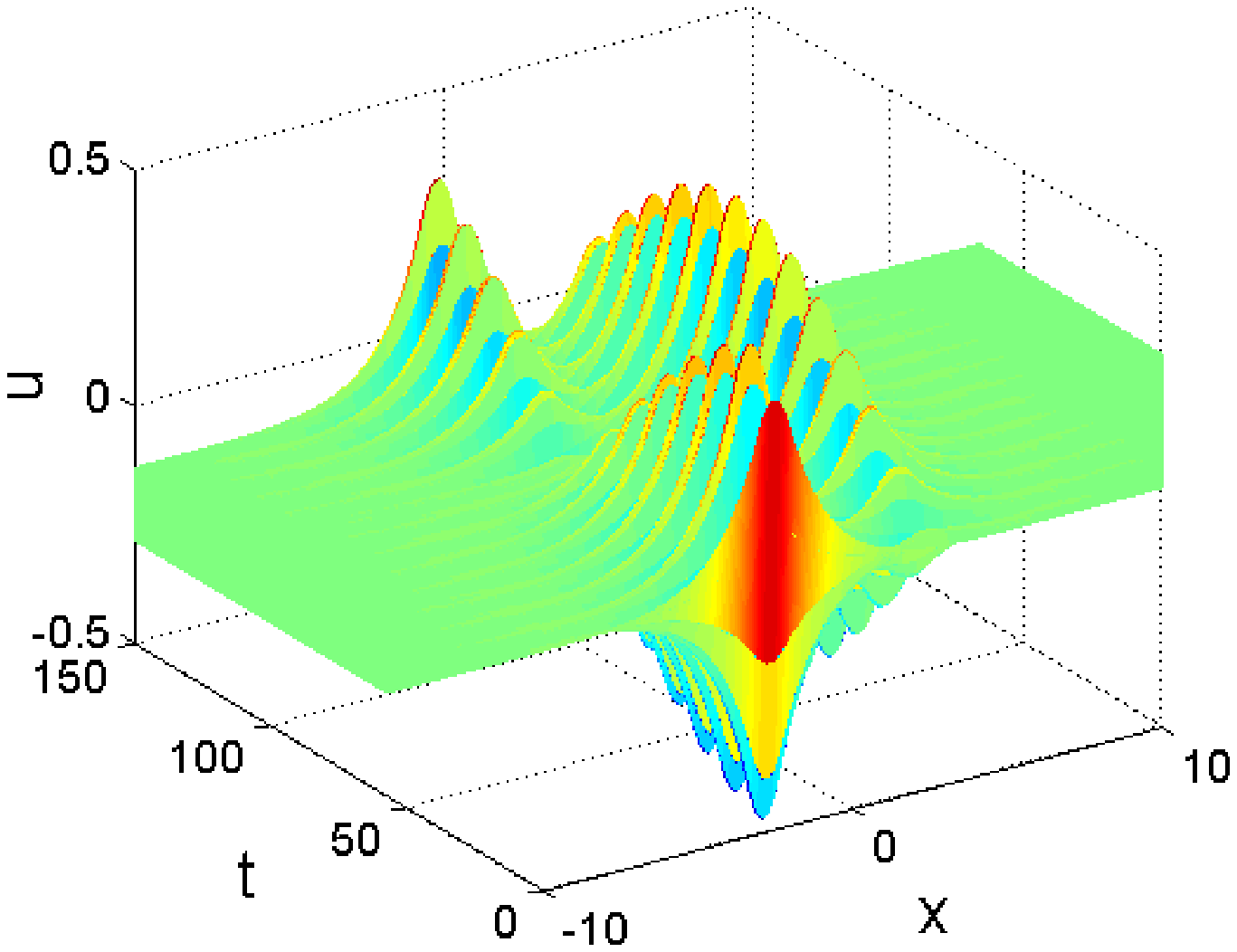}}
\subfigure[]{\includegraphics[width=4cm,angle=0,clip]{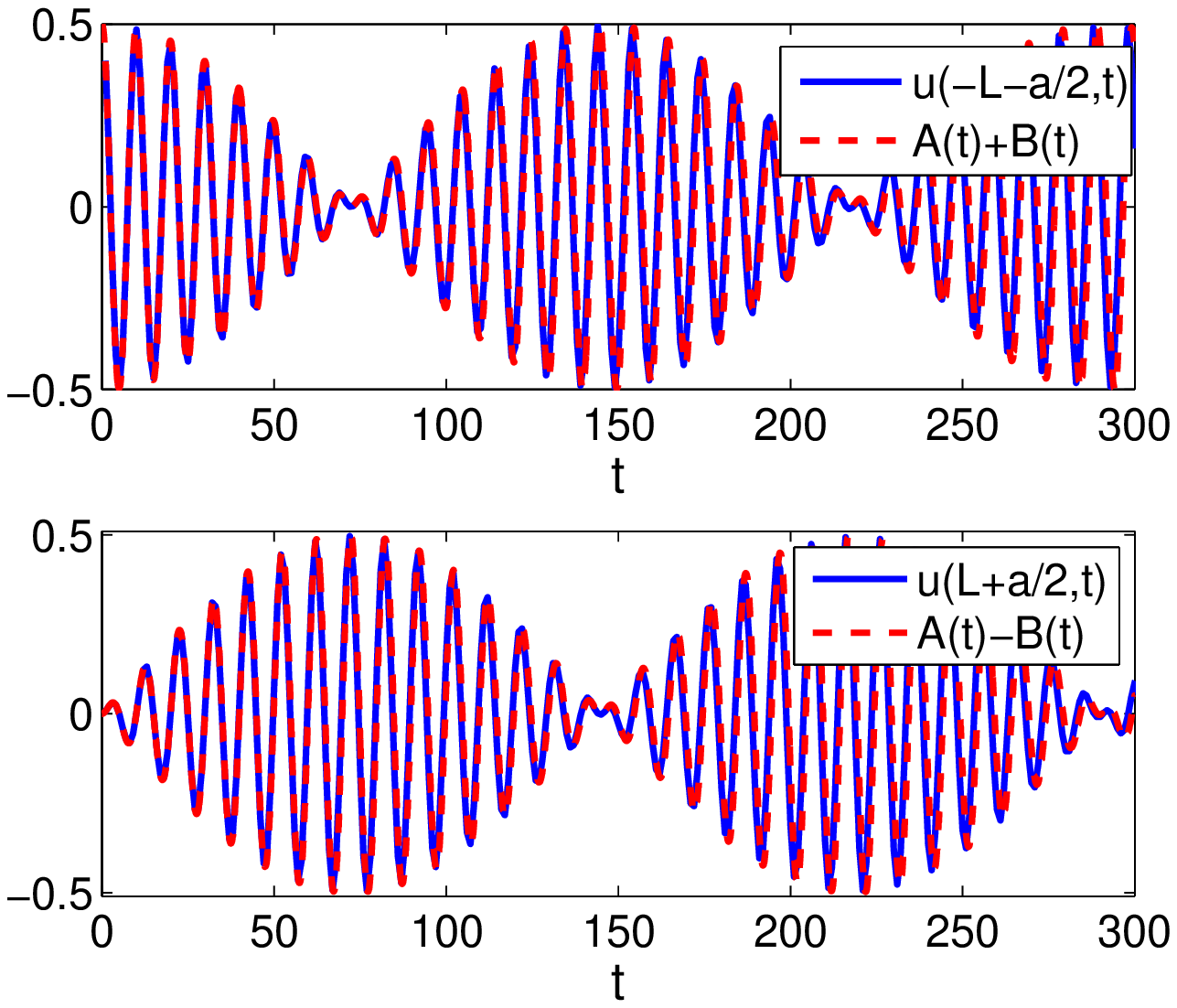}}\\
\subfigure[]{\includegraphics[width=4cm,angle=0,clip]{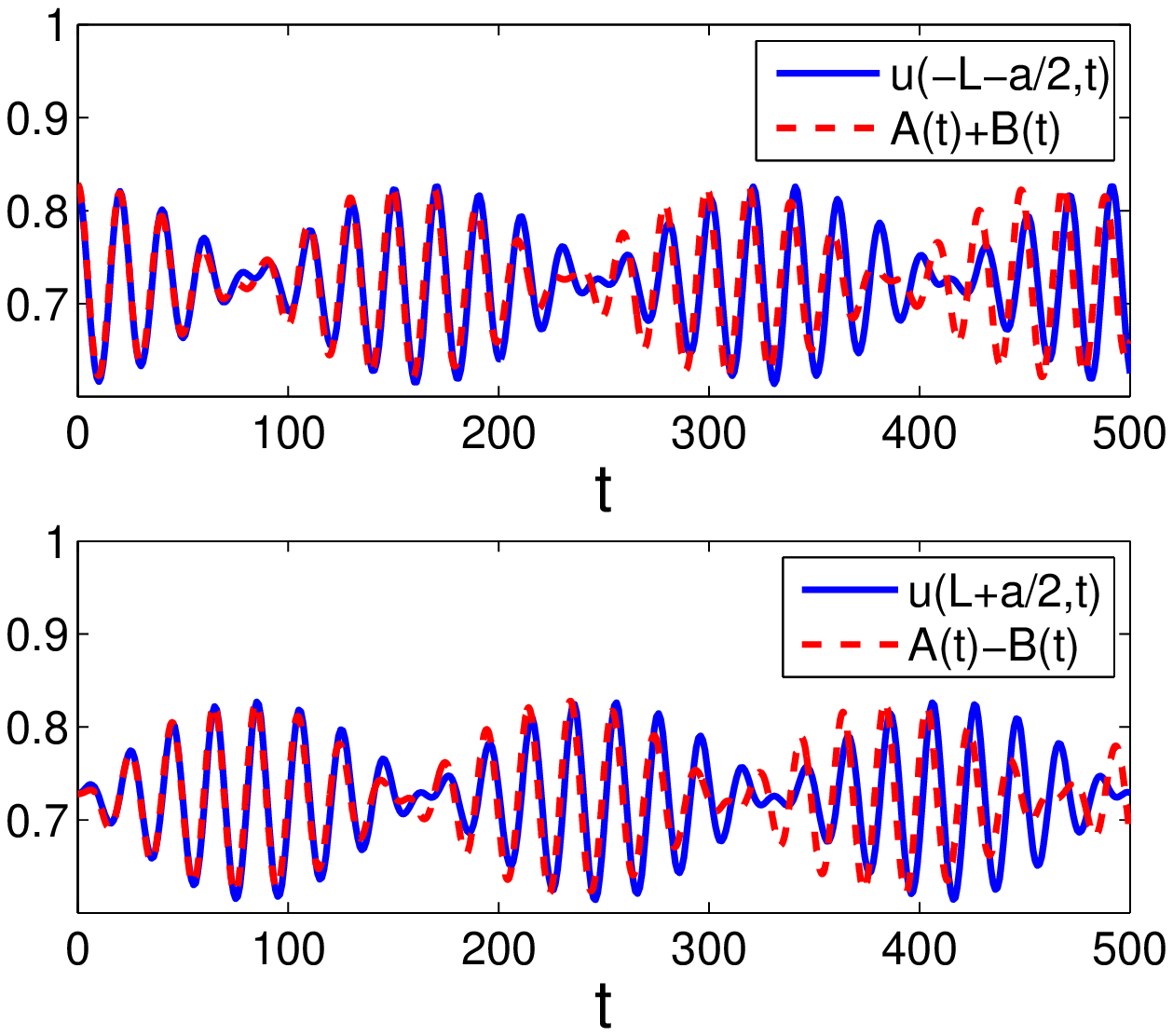}}
\subfigure[]{\includegraphics[width=4cm,angle=0,clip]{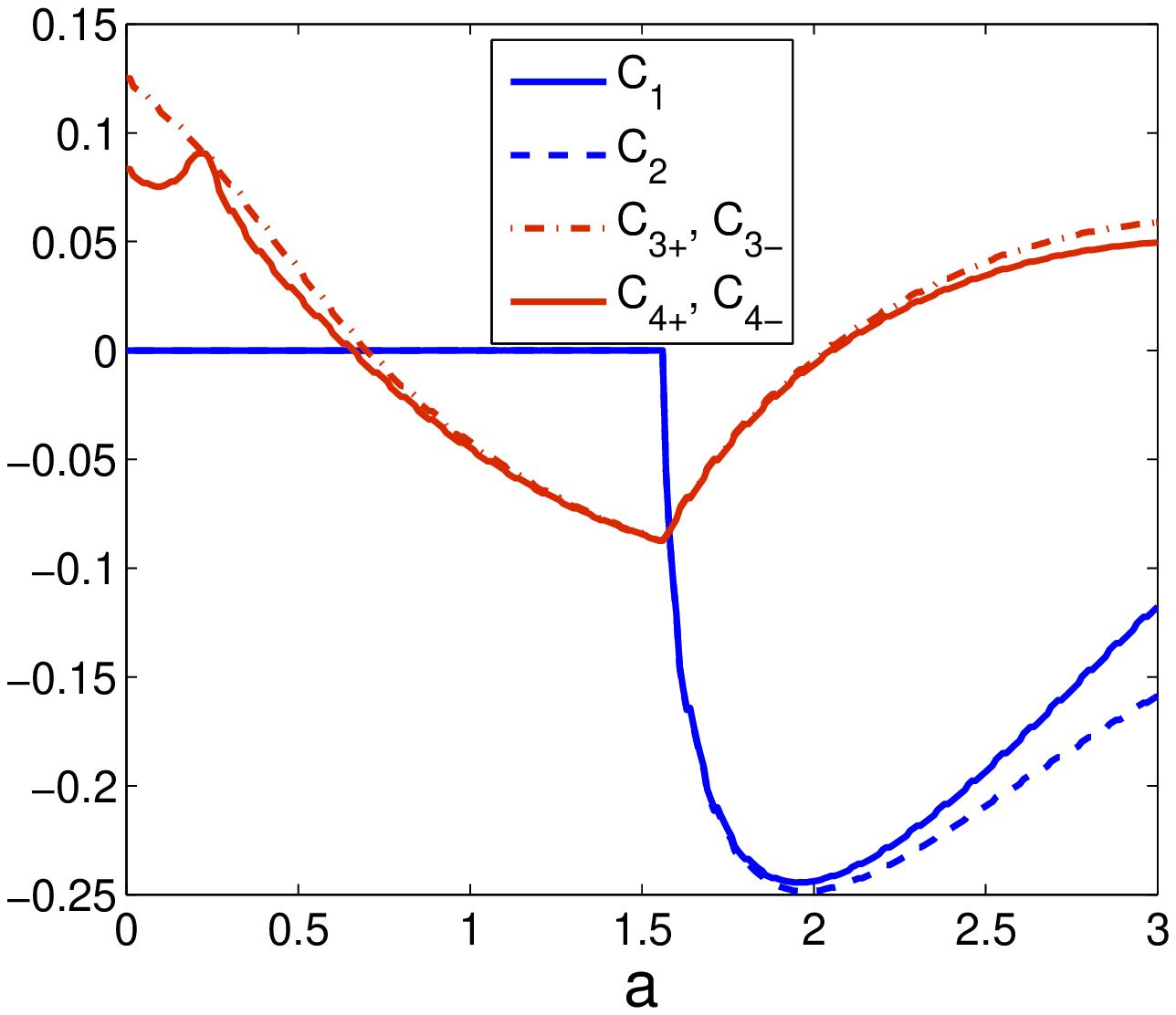}}
\caption{(Color online) Panel (a) shows the time dynamics of
  an initially localized excitation in the left well
 with $L=2$ and $a=1$.
 In panel (b), the oscillation amplitudes of the dynamics
 presented in panel (a) are plotted against time. Panel (c)
 depicts the oscillation amplitude on top of a nonzero
 background with a small initial amplitude
 for a double well with $L=2$ and $a=1.65$.
 Coupling coefficients appearing in (\ref{2mode}) as functions of $a$ for $L=2$ are shown in panel (d). In calculating the coefficients, the eigenfunctions have been normalized to $||\Phi_\pm||_\infty=1.$
 Approximations obtained from the two-mode equation (\ref{2mode}) are shown as red dashed curves.}
\label{fig1}
\end{figure}

\emph{Mode tunneling.} In the following, let us first consider
$L=2a=2$.  For those parameter values, $u=0$ is a stable ground state.
The numerically obtained time dynamics of an initially localized excitation in the left
well is presented in the top panels of Fig.\ \ref{fig1}, clearly
showing mode tunneling. Compare it with the time dynamics of BECs reported in Refs.\
\onlinecite{smer97,giov00,albi05,lebl10}.

Next, we consider a parameter combination of $L=2$ and $a=1.65$, i.e.\
$u=0$ is unstable. In the instability region of the constant solution,
excitations will oscillate on a non-zero background $u_{nc}$
(\ref{nc}). The oscillation amplitude in both wells as a function of
time is presented in panel (c) of Fig.\ \ref{fig1}. When the
initial oscillation amplitude is large enough, we interestingly obtain
chaotic oscillations (not shown here).

\emph{Two-mode approximations.} We will explain the observed mode
tunneling using a two-mode approximation. Looking for the solution of
the time-dependent equation (\ref{sg}) of the form
\begin{equation}
u(x,t)=u_{gs}+A(t)\Phi_++B(t)\Phi_-,
\label{ansatz}
\end{equation}
where $u_{gs}$ is the ground state of the system, i.e.\ $u_{gs}=0$ and
$u_{gs}=u_{nc}$ when $a$ is respectively on the left and right of the
vertical line in Fig.\ \ref{stabof0}, substituting the ansatz
(\ref{ansatz}) into (\ref{sg}), and projecting the equation onto
$\Phi_\pm$ will yield up to $\mathcal{O}(A^nB^{4-n}),\,n=0,\dots,4$
\begin{eqnarray}
\displaystyle
\begin{array}{lll}
\ddot{A} &=& \Lambda_+ A + \left(C_1A^2+C_{2}B^2\right) +
\left(C_{3+}A^3+C_{4+}AB^2\right); \\
\ddot{B} &=& \Lambda_- B + \left(C_{3-}B^3+C_{4-}A^2B\right);
\end{array}
\label{2mode}
\end{eqnarray}
with
\begin{align}
C_1=\frac{K_+}{2}\,\intd \theta\sin(u_{gs})\Phi_+^3\,dx,\\
C_{2}=\frac{K_+}{2}\,\intd \theta\sin(u_{gs})\Phi_+\Phi_-^2\,dx,\\
C_{3\pm}=\frac{K_\pm}{6}\,\intd \theta\cos(u_{gs})\Phi_\pm^4\,dx,\\
C_{4\pm}=\frac{K_\pm}{6}\,\intd \theta\cos(u_{gs})\Phi_+^2\Phi_-^2\,dx,
\end{align}
and $K_\pm = \left(\intd\Phi_\pm^2dx\right)^{-1}$. The constants $C_j$
are plotted against $a$ in panel (d) of Fig.\ \ref{fig1}. The
internal oscillation amplitude $u\left(\pm(L+a/2),t\right)$ is
respectively approximated by $u_{gs}+(A(t)\pm B(t))/2$.

For the uniform ground state ($u_{gs}=0$), we have solved
(\ref{2mode}) numerically and compared it with the oscillation
amplitude of the original equation in panel (b) of Fig.\
\ref{fig1}. One can observe that quantitative agreements are obtained.
An agreement is also obtained for mode tunneling on a non-uniform
background, as shown in panel (c) of Fig.\ \ref{fig1},
provided that the tunneling mode amplitude is small enough.

When $A(0)$ and $B(0)$ are large,
it is interesting to note that even though our two-mode approximation
does not quantitatively capture the dynamics of the chaotic tunnelings
it captures the qualitative transition to chaotic behavior.

\begin{figure}[tbhp]
\centering
\includegraphics[width=4cm,angle=0,clip]{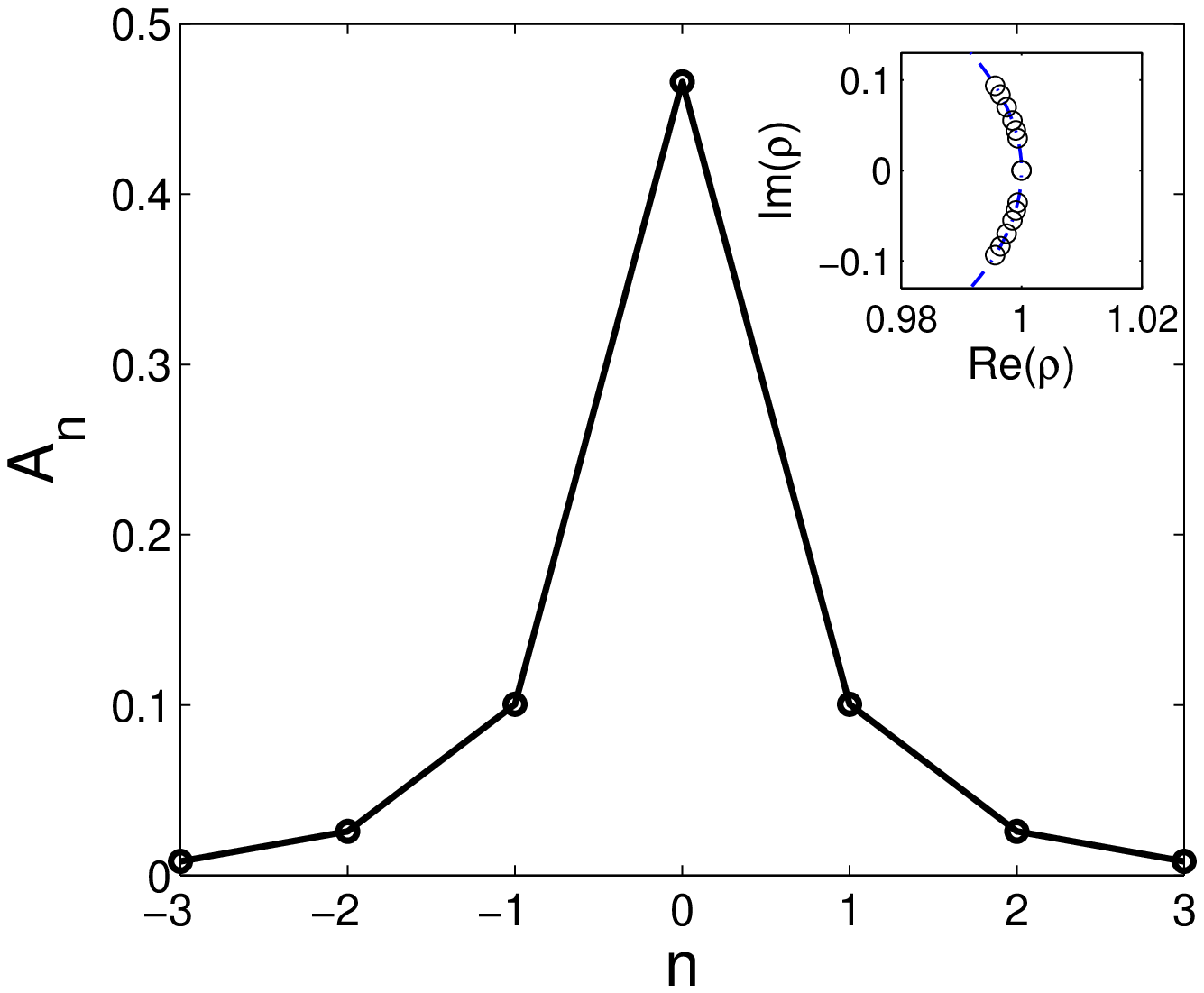}
\includegraphics[width=4cm,angle=0,clip]{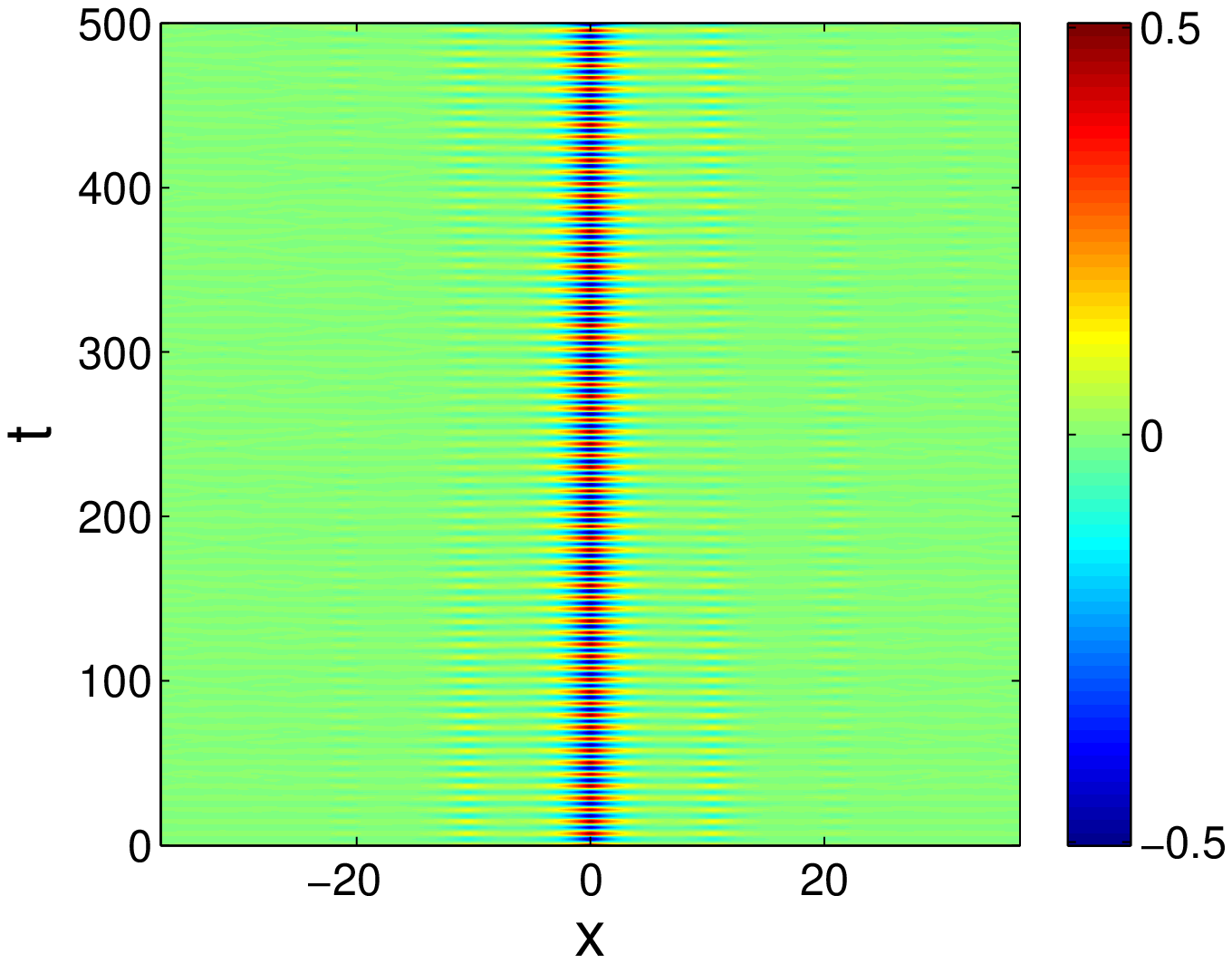}\\
\includegraphics[width=4cm,angle=0,clip]{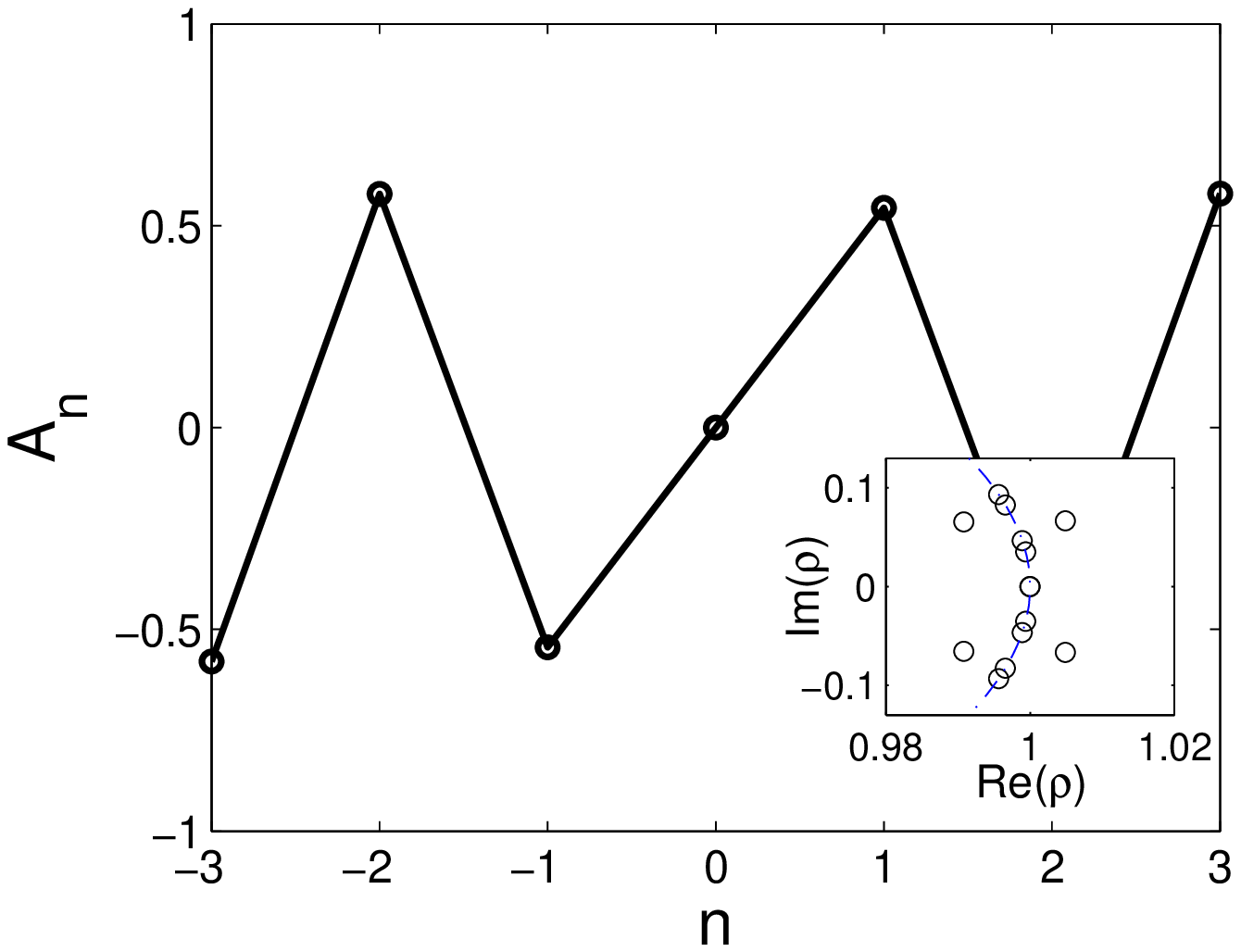}
\includegraphics[width=4cm,angle=0,clip]{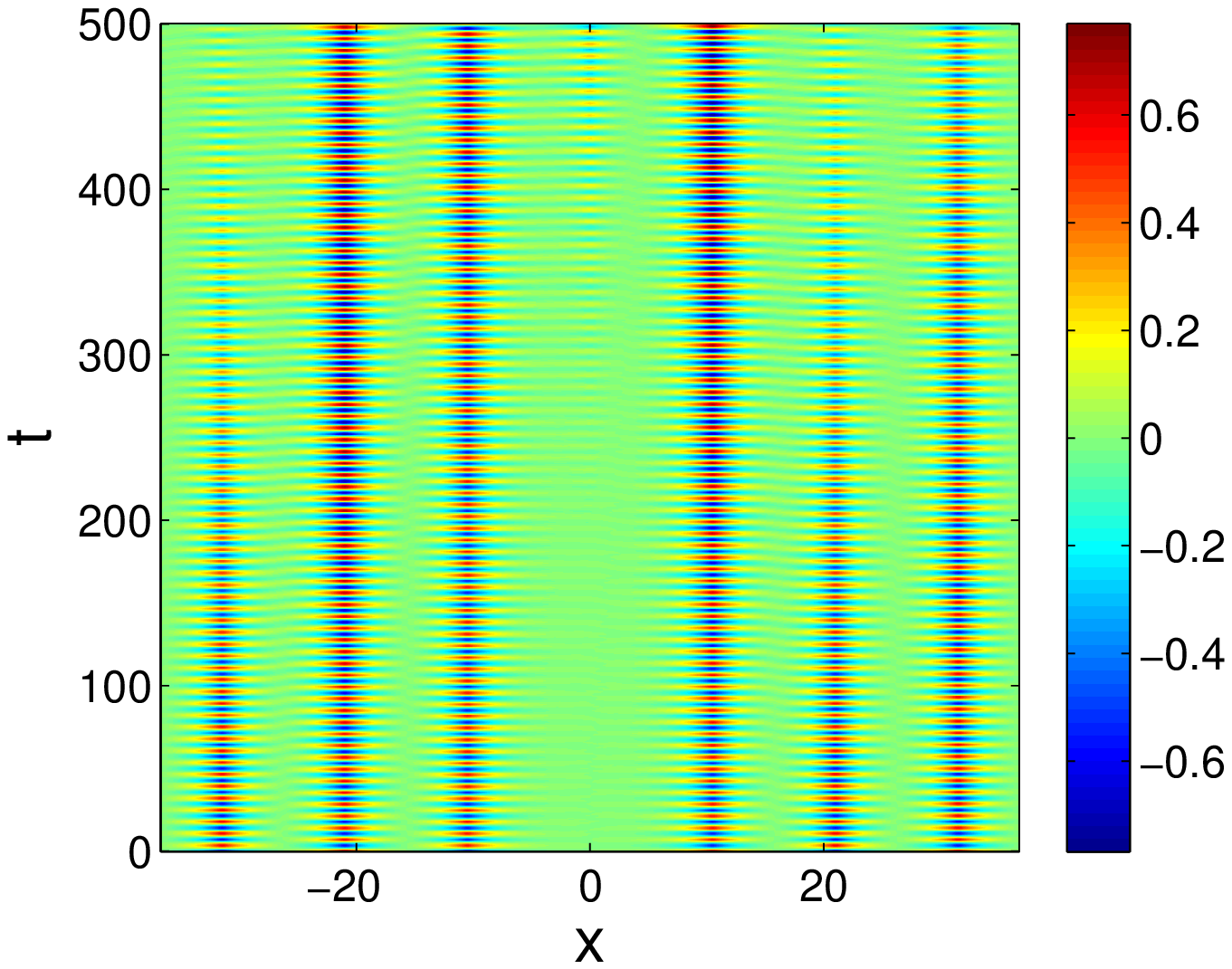}\\
\includegraphics[width=4cm,angle=0,clip]{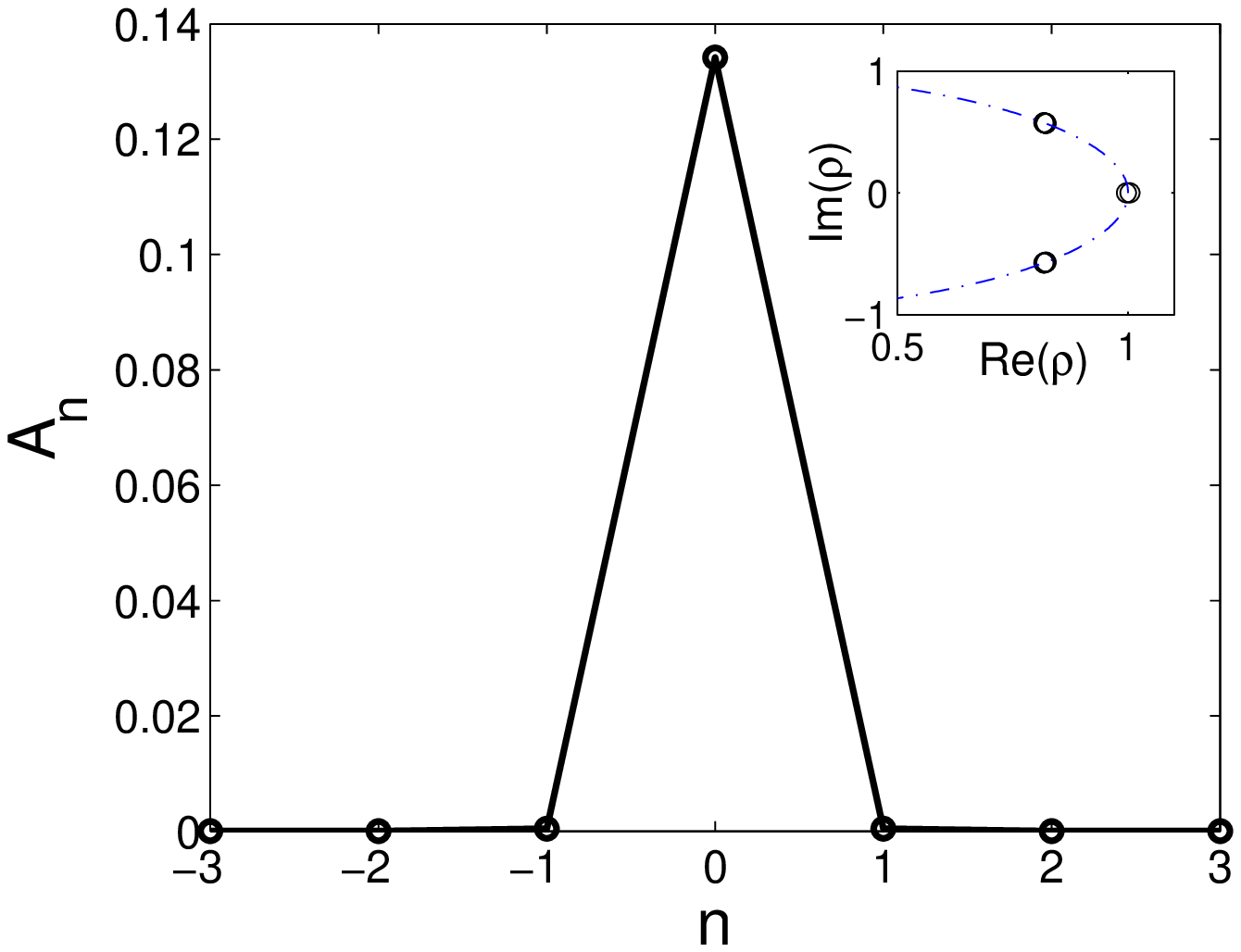}
\includegraphics[width=4cm,angle=0,clip]{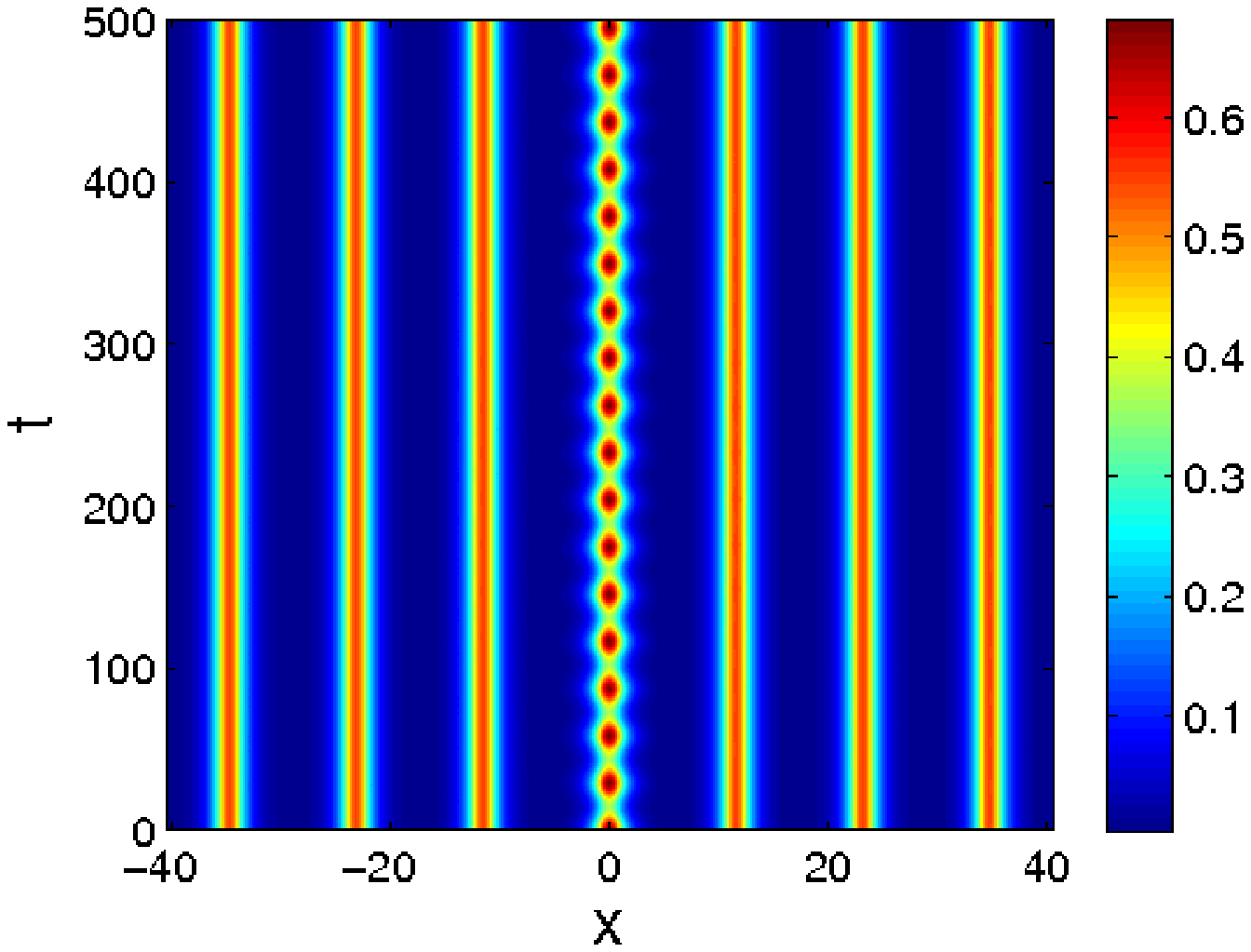}\\
\includegraphics[width=4cm,angle=0,clip]{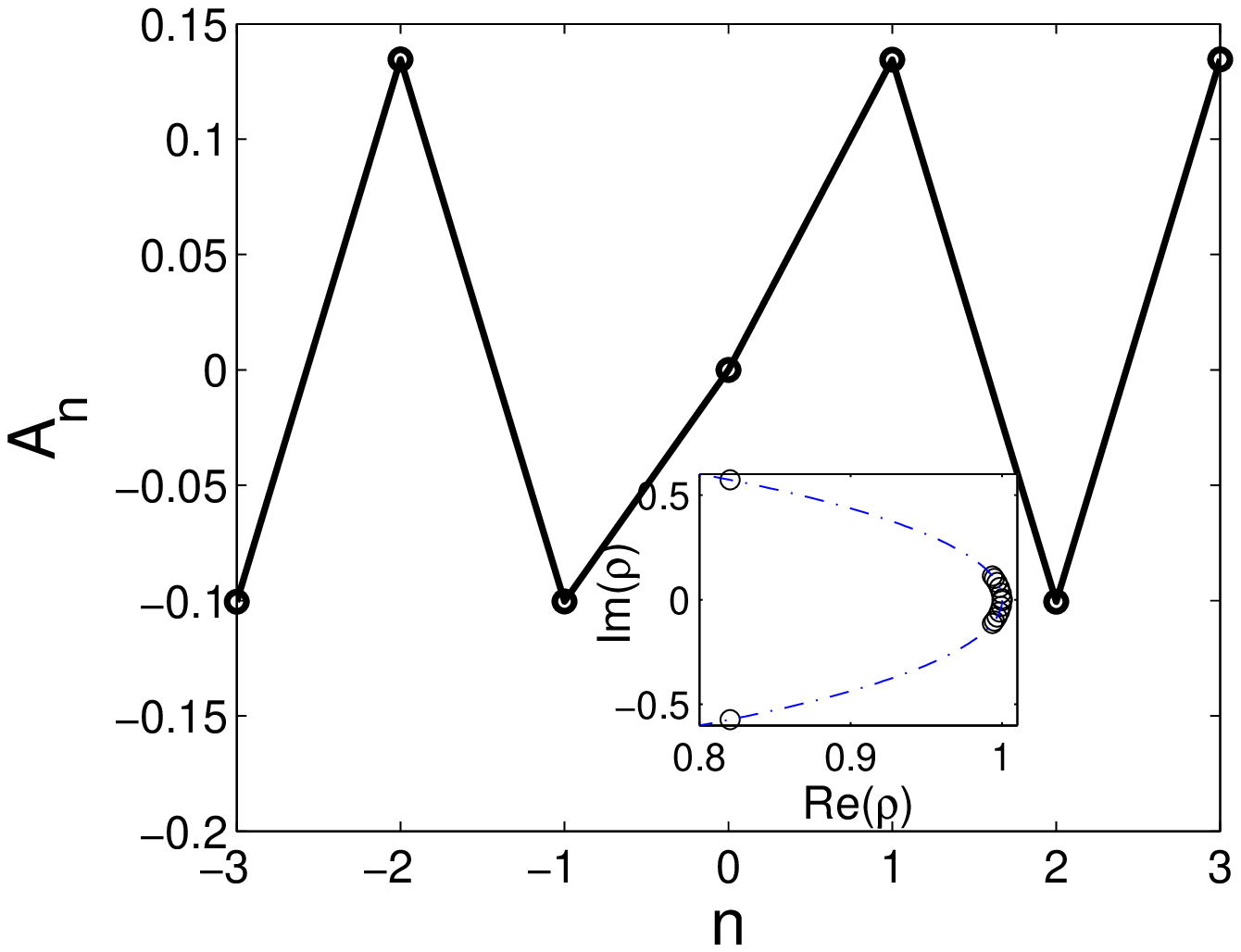}
\includegraphics[width=4cm,angle=0,clip]{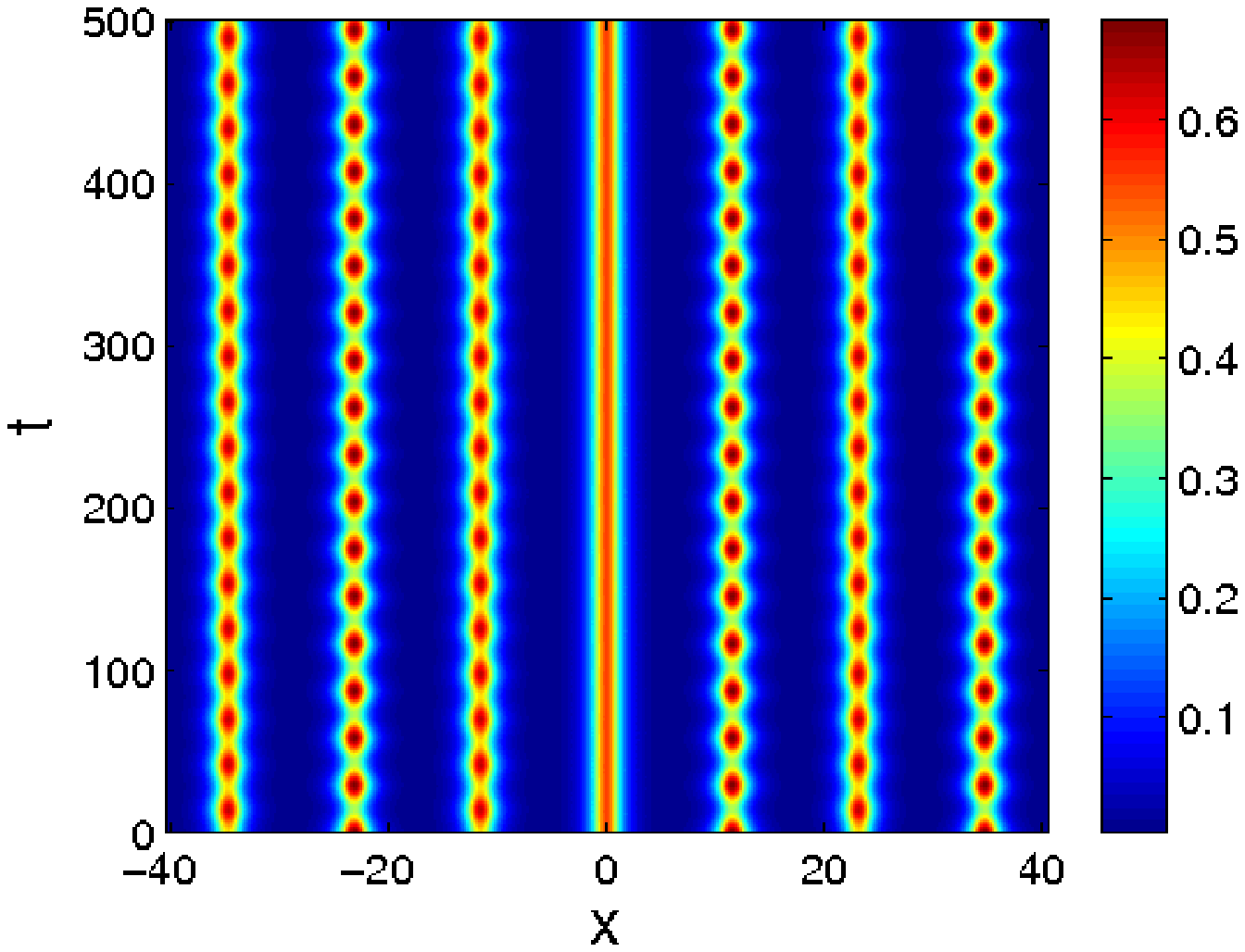}
\caption{(Color online) The left panels show the profile of
  numerically exact bright and dark lattice solitons obtained from the
  lattice equation (\ref{latt}). The insets depict the Floquet
  multipliers of the solution. The right panels present the
  corresponding time dynamics of (\ref{sg}) using the initial
  condition (\ref{ans}) with $A_n(0)$ shown in the adjacent left
  panel.
All panels have $L=10$, the upper two have $a=0.5$, the lower $a=1.6$.}
\label{fig3}
\end{figure}

\emph{Periodic defects.} One can also include more phase shifts and
derive a multi-mode approximation. Below we consider
the case of periodic shifts alternating $\pi$-junctions of length $a$ with
$0$-junctions of length $L$, i.e.,
\begin{equation}
\theta=\left\{
\begin{array}{ll}
-1, & x\in k(L+a)+(-\frac a2,\frac a2),k\in\mathbb{Z}\\
1, & \rm{elsewhere}.
\end{array}
\right.
\label{the2}
\end{equation}
It is known that in the limit $L\to\infty$ (i.e., one well), $u=0$ is
stable for $a<\pi/4$ and unstable otherwise \cite{susa03}.
The eigenfunction corresponding to the critical eigenvalue
$\Lambda_0$ of the ground state $u_{gs}$ in the limit $L\to\infty$
will be denoted by $\Phi_0(x)$. For $a<\pi/4$ and
$u_{gs}=0$,
\begin{equation}
\Phi_0(x)=\left\{
\begin{array}{ll}
\cos(\sqrt{1-\Lambda_0}a)
e^{-\sqrt{1+\Lambda_0}\left(|x|-a/2\right)},&|x|>a/2,\\
\cos(\sqrt{1-\Lambda_0}x),\,|x|<a/2.
\end{array}
\right.
\label{phi0}
\end{equation}

When $L\gg1$, a tight-binding approximation can be used to describe
the interaction of the defect modes in the system with periodic
defects. We write
\begin{equation}
u(x,t)=u_{gs}+\sum_{n=-\infty}^\infty A_n(t)\Phi_n(x),
\label{ans}
\end{equation}
where
$\Phi_n(x)=\Phi_0\left(x-n(a+L)\right)$. Performing the same procedure
as before, one will obtain the lattice equation
\begin{eqnarray}
&&\ddot{A}_n=\Lambda_0A_n+K_{2}A_n^2+K_{3}A_n^3\nonumber\\
&&+\!\sum_{j=\pm1}\!\!\!C_j[(K_{1,j}-\Lambda_0)A_{n+j}-K_{2}A_{n+j}^2-K_{3}A_{n+j}^3],
\label{latt}
\end{eqnarray}
where
\begin{eqnarray}
C_{\pm1}&=&K^{(0)}\int_{-\infty}^\infty\ \Phi_0(x)\Phi_{\pm1}(x)\,dx,\nonumber\\
K_{1,\pm1}&=&C_\pm^{-1}K^{(0)}\int_{-\infty}^\infty\Phi_0(x)\left(\partial_x^2-\theta\cos{u_{gs}}\right)\Phi_{\pm1}(x)\,dx,\nonumber\\
K_{2}&=&K^{(0)}\int_{-\infty}^\infty\frac12\sin(u_{gs})\Phi_0^3(x)\,dx,\nonumber\\
K_{3}&=&K^{(0)}\int_{-\infty}^\infty\frac16\cos(u_{gs})\Phi_0^4(x)\,dx,\nonumber
\end{eqnarray}
and $K^{(0)}=\left(\int_{-\infty}^\infty\Phi_0^2\,dx\right)^{-1}.$
Neglecting the nonlinear couplings to the neighboring sites ($CK_2,\,CK_3$),
the discrete equation above
becomes the lattice equation considered by Kivshar\cite{kivs93},
admitting many types of localized excitations. In the following, we
consider the special type, namely unstaggered bright and staggered
dark lattice solitons. In particular, we will show that the lattice
equation (\ref{latt}) can predict the stability of the solitons in the
original equation (\ref{sg}), provided that $L$ is large enough. In
doing so, we first solve Eq.\ (\ref{latt}) numerically for localized
modes using a shooting method and correspondingly study their
stability (see, e.g., the review Ref.\ \onlinecite{flac04}) and then
use the ansatz (\ref{ans}) at $t=0$ as an initial condition for the
governing equation (\ref{sg}). In the following, periodic boundary
conditions are used, which are relevant experimentally.\cite{gurl10}
Examples are shown in Fig.\ \ref{fig3}. Presented in the left and
right panels are numerically exact solutions obtained from the lattice
equation (\ref{latt}) and their corresponding time evolution in the
original equation (\ref{sg}).
The insets on the left panels depict the corresponding Floquet
multipliers, where the instability is indicated by the presence of
eigenvalues lying outside the unit circle (dash-dotted line).

First, we consider the parameter values $L=10$ and $a=0.5$,
representing the case of stable constant solution $u=0$. Shown in the
first and the second row are numerically exact bright and dark
solitons with the oscillation period $P=7.08$ ($\Lambda_0\approx-0.8$) and their dynamics.
According to the lattice equation (\ref{latt}),
the bright soliton is stable and the dark one unstable. One can note
from the right panels that the prediction provided by the lattice is
in agreement with the dynamics in the original system. The instability
of the dark soliton manifests in the form of the destruction of the
configuration.

Next, we consider the parameter values $L=10$ and $a=1.6$, which
represent the case of nonuniform ground state. The localized mode will
then oscillate on a nonzero background. Shown in the third and fourth
row are numerically exact bright and dark lattice solitons with the
oscillation period $P=30$ ($\Lambda_0\approx-0.0527$).

According to the lattice equation (\ref{latt}), the bright soliton has
the same stability as the case on stable uniform ground state, which
is confirmed by the time dynamics of the original equation. The
stability of the soliton for the chosen parameter values is not
surprising as the sites are rather uncoupled.

As for the dark soliton, it is interesting to note that in the present case 
it is stable, which is also confirmed by the dynamics of the full
equation. This implies that a nonzero background may act as a
stabilizer. Moreover, it is also important to note that the modes in
different lattices have different oscillation frequencies. The
multi-frequency breathers discussed in Ref.\
\onlinecite{kouk02,kouk04} may therefore be potentially observed in
experiments.

\emph{Conclusions.} We have considered Josephson junctions with
phase-shifts of $\pi$. By exploiting the defect modes present due to
the phase-discontinuities, the system has been shown to be an ideal
setting for studying mode interactions. In particular, we have shown
that mode tunneling in a double-well potential can be implemented in
the system
and presented the existence and stability of bright and dark solitons
in a periodic potential. We have shown that the analysis proved by a
multi-mode approximation gives a quantitative agreement with
dynamics of the original system.


\end{document}